\documentstyle[aps,prl,multicol,epsf]{revtex}
\title{Parity and Ruin in a Stochastic Game}
\author{E.~Ben-Naim$\dag$  and P.~L.~Krapivsky$\ddag$} 
\address{$\dag$Theoretical Division and Center for Nonlinear Studies, 
Los Alamos National Laboratory, Los Alamos, NM 87545, USA} 
\address{$\ddag$Center for Polymer Studies
  and Department of Physics, Boston University, Boston, MA 02215, USA}
\begin{document}
\maketitle
\begin{abstract}
\noindent     

We study an elementary two-player card game where in each round
players compare cards and the holder of the smallest card wins.  Using
the rate equations approach, we treat the stochastic version of the
game in which cards are drawn randomly.  We obtain an exact solution
for arbitrary initial conditions.  In general, the game approaches a 
steady state where the card densities of the two players are
proportional to each other. The leading small size behavior of the
initial card densities determines the corresponding proportionality
constant, while the next correction governs the asymptotic
time dependence.  The relaxation towards the steady state exhibits a
rich behavior, e.g., it may be algebraically slow or exponentially
fast.  Moreover, in ruin situations where one player eventually wins
all cards, the game may even end in a finite time.

\medskip\noindent

{PACS numbers: 02.50.Ey, 05.40.-a, 51.10.+y, 89.65.-s}
\end{abstract}
       
\begin{multicols}{2}
   
\noindent

Numerous phenomena in social and economic sciences involve multiple
interacting agents. The interaction between these agents often leads
to exchange of quantities such as capital, goods, political opinions,
etc.~\cite{lig,soc,ikr,CMV,fkb}.  Games are widely employed in modelling
collective behavior especially in the context of economics\cite{vN},
with recent examples ranging from evolution of trading strategies in a
stock market \cite{cz,cmz,c} to bidding in auctions \cite{dr}.  Games
can often be regarded as many body exchange processes resembling
collision processes \cite{col}, and therefore their dynamics may be
described by suitably adapted kinetic theories \cite{rl}. Here, we
investigate a stochastic null strategy card game.  By considering the
``thermodynamic limit'' where the initial number of cards is infinite,
we show that rate equations provide a natural framework for analyzing
game dynamics.

Our two-player game is defined as follows.  Each player starts with a certain
number of cards.  At each round players draw a card randomly from their deck
and compare the card values. The player with the smaller card wins the round
and gets both cards.  This is repeated ad infinitum or until one of the
players gains all cards.  This stochastic adaptation of the elementary card
game ``war'' is also motivated by a recently introduced auction bidding model
where the lowest bid is rewarded \cite{dr}.

Our main result is that one specific aspect of the initial card
distribution, namely, the small size tail, governs the dynamics of the
game. Let us denote by $A$ and $B$ the two players, and let their
initial card densities be $a_0(x)$ and $b_0(x)$, respectively. In the
long time limit, a  steady state is approached with the card
densities of both players being equal to a fraction of the total card
density, $a_\infty(x)=\alpha [a_0(x)+b_0(x)]$, with $a_\infty(x)$ the
limiting card density of player $A$. While a family of steady state
solutions characterized by the parameter $0\leq\alpha\leq 1$ is in
principle possible, the leading small size behavior of the initial
distributions selects a specific value $\alpha=\lim_{x\to
  0}{a_0(x)\over a_0(x)+b_0(x)}$.  Moreover, the next leading
correction determines how the system approaches the steady state. The
corresponding time dependent behavior may be algebraic or exponential.
Interesting behaviors also occur when one player captures all cards.
In this case, the game duration maybe finite or infinite.
Additionally, using numerical simulations we show that the theoretical
predictions concerning the game duration extend to deterministic
realizations of the game.

Let the initial number of cards of player $A$ and $B$ be $N_A$ and $N_B$,
respectively, and let the total number be $N=N_A+N_B$.  We shall take the
thermodynamic limit $N_A,N_B,N\to\infty$ such that the initial fractions
$N_A/N$ and $N_B/N$ are fixed.  Then the card exchange process can be
conveniently described via rate equations for $a(x,t)$ and $b(x,t)$, the
densities of cards with value $x$ at time $t$ for players $A$ and $B$,
respectively.  These densities evolve according to the nonlinear
integro-differential equations
\begin{equation}
\label{rate}
{\partial\over\partial t} a(x,t)=R(x,t), \qquad
{\partial\over\partial t} b(x,t)=-R(x,t),
\end{equation}
with the gain (loss) term $R(x,t)$ given by
\begin{eqnarray*}
R={1\over A(t)B(t)}
\left[b(x,t)\!\int_0^x\!\! dy\,a(y,t)-a(x,t)\!\int_0^x\!\! dy\, 
b(y,t)\right].
\end{eqnarray*}
Here  
\begin{equation}
\label{ab}
A(t)=\int_0^\infty dx\, a(x,t), \qquad
B(t)=\int_0^\infty dx\, b(x,t)
\end{equation}
are the fraction of cards possessed by players $A$ and $B$,
respectively.  Clearly, 
\begin{equation}
\label{ab1}
A(t)+B(t)=1.
\end{equation}
The rate equations (\ref{rate}) reflect the nature of the game as the
rate by which player A gains (loses) cards of value $x$ is proportional to
the fraction of his opponent's cards which is larger (smaller) than $x$.  The
overall factor $[AB]^{-1}$ ensures that on average, every opposing pair of
cards comes into play once per unit time. The minimal card value was tacitly
set to zero as the process is invariant under the transformation $x\to x +
{\rm const}$.

Besides the obvious conservation law (\ref{ab1}), two other conservation laws
underly the process. First, the total number of cards of a given value is
conserved,
\begin{equation}
\label{cons} 
a(x,t)+b(x,t)=u_0(x),
\end{equation}
where $u_0(x)=a_0(x)+b_0(x)$ is the initial total density. Second, the
density of the minimal card remains constant throughout the evolution:
$a(0,t)=a_0(0)$, where $a_0(x)\equiv a(x,t=0)$, and similarly for $B$.

The steady state behavior is quite generic. The cumulative card
densities, ${\cal A}(x,t)=\int_0^x dy\, a(y,t)$ and ${\cal
  B}(x,t)=\int_0^x dy\, b(y,t)$, satisfy ${\cal A}'/{\cal A}={\cal
  B}'/{\cal B}$ in the long time limit.  Thus, ${\cal
  A}_\infty(x)\propto {\cal B}_\infty(x)$, and consequently, the
limiting card densities, $a_\infty(x)={\cal A}'_\infty(x)$ and
$b_\infty(x)={\cal B}'_\infty(x)$, are proportional to each other.  The
conservation law (\ref{cons}) implies that each of the limiting card
densities equals a fraction of the overall card density
\begin{eqnarray}
\label{final}
a_\infty(x)=\alpha u_0(x),\qquad 
b_\infty(x)=(1-\alpha) u_0(x).
\end{eqnarray}

In principle, for a given total card density $u_0(x)$, there is a family of
steady state solutions characterized by \hbox{$0\leq\alpha\leq 1$}. Moreover,
initial conditions where the densities are proportional to each other do not
evolve further in time regardless of $\alpha$. Still, for a given initial
condition a specific value of $\alpha$ is selected.  This value can easily be
found for a class of initial conditions with non-vanishing minimal card
densities, $u_0(0)>0$. Consider the density of the smallest cards $x=0$.
Equation (\ref{final}) gives $a_\infty(0)=\alpha u_0(0)$, while the second
conservation law implies $a_\infty(0)=a_0(0)$, and hence
$\alpha=a_0(0)/[a_0(0)+b_0(0)]$. This simple argument suggests that the
smallest cards govern the outcome of the game. In the following, we solve for
the full time dependent behavior and show that in general, the limiting small
size tail of the two distributions dictates $\alpha$.

To solve the time dependent behavior, we make two simplifying
transformations.  First, the overall rate by which the exchange occurs
$[AB]^{-1}$ can be absorbed into a modified time variable $\tau$, defined via
\begin{equation}
\label{tau}
\tau=\int_0^t ds\, \left[A(s)B(s)\right]^{-1}.
\end{equation}
The second transformation essentially reduces any total density
$u_0(x)$ to a uniform density by introducing the variable $\xi$
\begin{equation}
\label{xi} 
\xi=\int_0^x dy\,u_0(y).
\end{equation} 
Suppressing the explicit time dependence, the transformed card
densities are found from the relations $\bar{a}(\xi)\,d\xi=a(x)\,dx$
and $\bar{b}(\xi)\,d\xi=b(x)\,dx$.  Clearly, these satisfy
$\bar{a}(\xi)=a(x)/u_0(x)$ and $\bar{b}(\xi)=b(x)/u_0(x)$.  In the
following, we shall omit the bar.  The conservation law (\ref{cons}) 
becomes 
\begin{equation}
\label{norm} 
a(\xi,\tau)+b(\xi,\tau)=1,
\end{equation}
i.e., the transformed total density is uniform on the interval [0,1] (note
that Eqs.~(\ref{ab1}) and (\ref{xi}) imply $0\leq \xi\leq 1$).

The above transformations simplify the evolution equations, and given
the linear dependence (\ref{norm}), it suffices to solve for $a$
\begin{eqnarray*}
{\partial \over \partial \tau}a(\xi,\tau)
=b(\xi,\tau)\!\int_0^\xi\!\! d\eta\,a(\eta,\tau)-a(\xi,\tau)\!
\int_0^\xi\!\! d\eta\,b(\eta,\tau).
\end{eqnarray*}
Replacing $b(\xi,\tau)$ with $1-a(\xi,\tau)$ linearizes this equation 
\hbox{${\partial \over \partial \tau}a(\xi,\tau)=\int_0^\xi 
\!d\eta\,a(\eta,\tau)-\xi a(\xi,\tau)$},
and differentiating with respect to $\xi$ yields further simplification 
\begin{equation}
\left({\partial \over \partial \tau}+\xi\right)
{\partial \over \partial \xi}a(\xi,\tau)=0.
\end{equation}
Integrating over $\tau$ and then over $\xi$ we arrive at our primary
result, the exact time dependent solution for arbitrary initial
conditions: 
\begin{equation}
\label{solution}
a(\xi,\tau)=\alpha+\int_0^{\xi} d\eta\, a_0'(\eta)\,e^{-\eta\tau}.
\end{equation}
Hereinafter we utilize the notations $a_0(\xi)\equiv a(\xi,\tau=0)$,
$a_0'(\xi)\equiv {d \over d \xi}a_0(\xi)$, and $\alpha=a_0(\xi=0)$.

Let us again consider the steady state. In the long time limit
$\tau\to\infty$, the integral in (\ref{solution}) vanishes and the
densities become uniform $a(\xi,\tau)\to \alpha$ and $b(\xi,\tau)\to
1-\alpha$.  Hence in terms of the original variable $x$, both
densities are proportional to $u_0(x)$ according to Eq.~(\ref{final}),
with $\alpha=a_0(\xi=0)=a_0(x=0)/u_0(x=0)$.  Even when
$u_0(x)$ vanishes or diverges as $x\to 0$, the parameter $\alpha$ is
well-defined and using l'Hopital rule, it is given by
\begin{equation}
\label{alpha}
\alpha=\lim_{x\to 0} {a_0(x)\over a_0(x)+b_0(x)}.
\end{equation}
Thus, if the two distributions exhibit different leading behaviors, say
$\lim_{x\to 0} b_0(x)/a_0(x)=0$, then player $A$ eventually ruins player $B$.
Hence, the small card tail $x\to 0$ provides the necessary selection criteria
determining which of the family of solutions (\ref{final}) is eventually
selected by the dynamics.

We now study the approach to the steady state. For example, 
the density $A(\tau)=\int_0^1 d\xi\, a(\xi,\tau)$ is given by
\begin{equation}
\label{Atau} 
A(\tau)=\alpha+\int_0^1 d\xi\, (1-\xi)\,a_0'(\xi)\, e^{-\xi\tau}.
\end{equation}
While the steady state behavior is determined by the leading small
argument behavior of $a_0(\xi)$, the relaxation towards the final
state is governed by the correction to the leading behavior. Let us
consider the following small argument behavior
\begin{equation}
\label{dens}
a_0(\xi)\simeq\alpha+\gamma \xi^{\delta}\qquad \xi\to 0,
\end{equation}
with $\delta>0$.  Then, one has 
$A(\tau)-\alpha\simeq\gamma\Gamma(\delta+1)\tau^{-\delta}$.
However, in terms of the actual time variable $t$, a richer variety of
behaviors is exhibited.

First, suppose that the system approaches an active steady state,
i.e., $0<\alpha<1$.  Then from Eq.~(\ref{tau}) we obtain $t\to
\alpha(1-\alpha)\tau$, and therefore 
\begin{equation}
A(t)-\alpha\simeq Ct^{-\delta}, \qquad t\to \infty        
\end{equation}
with $C=\gamma\Gamma(\delta+1)[\alpha(1-\alpha)]^{\delta}$. Hence, the
approach is generally algebraic.

Next, suppose that one player, say $A$, eventually ruins their opponent, i.e.,
$\alpha=1$. Then $dt/d\tau\sim B(\tau)\sim \tau^{-\delta}$ and consequently,
$t\sim \tau^{1-\delta}$. Therefore, for $\delta\leq 1$ representing weak
initial advantage of the eventual winner, the game duration is infinite:
\begin{equation} 
1-A(t)\sim 
\cases{t^{-{\delta\over 1-\delta}}&$\delta<1;$\cr
e^{-{\rm const}\times t}&$\delta=1$.\cr}
\end{equation}
In the complementary situation of strong initial advantage for the
eventual winner, $\delta>1$, the game duration is finite:
\begin{equation}
\label{finite}
A(t_f)=1.
\end{equation}
The terminal time can be determined from the integral 
\hbox{$t_f=\int_0^\infty d\tau A(\tau)\left[1-A(\tau)\right]$}.  Using
Eq.~(\ref{Atau}) and recalling that $\alpha=1$ yields this time as an
explicit function of the initial conditions
\begin{eqnarray}
\label{tf}
t_f=&-&\int_0^1 d\xi\,{1-\xi\over \xi}\,a_0'(\xi)\\
&-&\int_0^1 \int_0^1 d\xi_1\,d\xi_2\, {(1-\xi_1)(1-\xi_2)\over \xi_1+\xi_2}\,
a_0'(\xi_1)\,a_0'(\xi_2).\nonumber
\end{eqnarray}
For example, the initial density $a_0(\xi)=1-\xi^2$ yields 
$t_f={2\over 15}+{16\over 15}\,\ln 2\approx 0.87269$. Additionally,
the time dependent approach towards the final state is algebraic, 
\begin{equation}
1-A(t)\sim (t_f-t)^{\delta\over\delta-1},
\end{equation}
sufficiently close to the terminal time $t\to t_f$. As expected, the density
decreases linearly with time when the disparity between the two players
becomes very large in the limit $\delta\to\infty$.

Thus if the system approaches a trivial steady state with one player winning
all cards, the temporal behavior can be algebraically slow or exponentially
fast.  Moreover, every positive power can be realized.  Remarkably, if the
initial disparity between the two players is sufficiently large, the game
ends in a finite time.  Interestingly, such disparity is expressed only in
terms of the density of the smallest cards, with the larger cards practically
irrelevant to the game outcome.

Next, we analyze the time dependent evolution of the entire card density, not
simply the overall number density.  Evaluating the leading behavior of the
density (\ref{solution}) in the long time limit, we find that the density
exhibits a boundary layer structure
\begin{equation}
a(\xi,\tau)-\alpha\simeq
\cases{\gamma\xi^\delta&$\xi\ll \tau^{-1}$;\cr
\gamma\Gamma(\delta+1)\tau^{-\delta}&$\xi\gg \tau^{-1}$.\cr}
\end{equation} 
The scale $\xi_0\sim \tau^{-1}$ underlies the
distribution. Cards larger than this scale have already relaxed to the
limiting distribution, while cards smaller than this scale have yet to
exchange hands and hence, are still dominated by the initial
distribution. In other words, smaller cards are slower to equilibrate,
consistent with the fact that the small size tail dominates the
asymptotic behavior.

We now briefly discuss the case where the number of card flavors is finite,
or in other words, discrete card distributions
\begin{eqnarray}
\label{disc}
a(x,t)&=&\sum_{n=1}^k a_n(t)\delta(x-x_n),\\   
b(x,t)&=&\sum_{n=1}^k b_n(t)\delta(x-x_n),\nonumber
\end{eqnarray}
with $x_1=0$ and $x_n<x_{n+1}$. The discrete version of the rate equations
can be written and solved directly using a series of transformation which
mimics the ones used above. Instead, we shall insert the
initial conditions (\ref{disc}) in the general continuous case solution
(\ref{solution}).

Denote again $u_n(t)=u_n(0)=a_n(0)+b_n(0)$ the total card concentration.  The
variable $\xi_n=\sum_{m=1}^{n-1} u_m(0)$ plays the role of $\xi$ and the time
variable $\tau$ remains as in Eq.~(\ref{tau}). The solution (\ref{solution})
reads
\begin{equation}
\label{sol-disc}
{a_n(\tau)\over u_n(0)}={a_1(0)\over u_1(0)}
+\sum_{m=2}^n\left({a_m(0)\over u_m(0)}-{a_{m-1}(0)\over u_{m-1}(0)}\right)
e^{-\xi_m\tau}.
\end{equation}
Since all terms in the summation eventually vanish, the two players approach
a limiting distribution which is proportional to the initial distribution
$a_n(\infty)=\alpha u_n(0)$ with $\alpha=a_1(0)/u_1(0)$, in accordance with
Eq.~(\ref{alpha}). In general, the approach to the steady state is
exponential. We first discuss the case $0<\alpha<1$.  Since $A_\infty=
\alpha$, one has $t\to
\alpha(1-\alpha)\tau$ asymptotically.  Hence, the relaxation towards the
steady state is exponential
\begin{equation}
A(t)-\alpha\sim e^{-{\rm const} \times t}.
\end{equation}
In the complementary case when one player wins all cards, $\alpha=1$, 
the approach is dominated by the first non-vanishing term in the
summation, namely, the first non-vanishing $b_n(0)$.  In this case,
$dt/d\tau\propto \exp(-{\rm const}\times \tau)$,
and consequently, the game duration is finite as in
Eq.~(\ref{finite}).  In summary, the behavior in the discrete case is
different from the continuous case in that the time dependent behavior
is generally exponential. An additional difference is that when one player
captures all cards, the game duration is always finite.

In the above, we discussed games with an infinite number of cards.
Nevertheless, one can apply the above results to  realistic situations
when both players start with a finite number, say $N$, cards.  We note that
the time unit used above is of the order $N^2$ rounds in an actual game.  For
the case $\delta>1$ one therefore predicts a duration
\begin{equation}
\label{diffuse}
T_f\sim N^2, 
\end{equation}
with $T_f$ the number of rounds. The duration in the marginal $\delta=1$ case
can be estimated using the average time it takes for player $B$ to get down
to one card $B(t)=N^{-1}$.  Utilizing the exponential decay of
$B(t)$, we find that there is an additional logarithmic dependence, $T_f\sim
N^2\ln N$, in this case.

\begin{figure}
\centerline{\epsfxsize=8cm \epsfbox{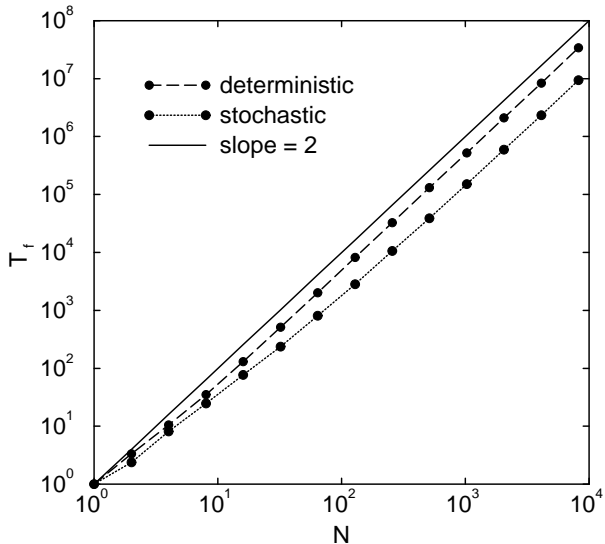}}
\caption{ Duration of the game as a function of the number of cards. 
Shown is $T_f$ the number of rounds a game lasts on average versus $N$, 
the initial number of cards.  The results represent an average over $10^4$
realizations. A line of slope $2$ is plotted for reference.}
\end{figure}

Monte Carlo simulations are consistent with these predictions. In the
simulations, each player starts with $N$ cards drawn from a uniform
distribution in the range $0<x<1$. Eventually, the player holding the
smallest card wins. Our theory describes the stochastic realization of the
game where cards are drawn randomly from the deck. We also examined the
deterministic case where the card order is fixed throughout the game.  In
this version, the winner of a round places both cards in the bottom of the
deck. In both cases, we find diffusive terminal times as in
Eq.~(\ref{diffuse}). Nevertheless, the two cases differ with the stochastic
game ending faster than the deterministic one (see Fig.~1).  Additionally, we
find that this diffusive time scale characterizes the entire distribution of
terminal times, as fluctuations in the terminal time are proportional to the
mean $\langle T_f^2\rangle-\langle T_f\rangle^2\propto \langle T_f\rangle^2$.

In closing, we studied a stochastic two-player card game using the
rate equations approach.  We found that extremal characteristics of
the initial conditions select a particular steady state out of a
family of possible solutions.  Eventually, the card densities of the
players become proportional to each other.  However, the players
generally possess different overall number of cards and it is even
possible that one player gains all cards.  The approach towards the
steady state exhibits rich behavior. Large cards tend to equilibrate
faster than small cards, and the distribution develops a boundary
layer structure. The time dependent behavior of the overall density is
algebraic in cases where a steady state is approached. In the
complementary case where one player gains all cards, the game may end
in a finite or an infinite time. The relative initial advantage of the
winner, characterized by the correction to the leading extremal
behavior, determines the game duration in this case.

This research was supported by DOE (W-7405-ENG-36) and NSF(DMR9978902).

\end{multicols}

\begin{references}

\bibitem{lig} T.~M.~Liggett, {\it Interacting Particle Systems}
              (Springer, New York, 1985).
  
\bibitem{soc} R.~Axelrod, {\it The Complexity of Cooperation}
              (Princeton University Press, Princeton, 1997).

\bibitem{ikr} S.~Ispolatov, P.~L.~Krapivsky, and S.~Redner, 
              Eur.\ Phys.\ J. B {\bf 2}, 267 (1998).

\bibitem{CMV} C.~Castellano, M.~Marsili, and A.~Vespignani, 
              Phys.\ Rev.\ Lett.\ {\bf 85}, 3536 (2000). 
 
\bibitem{fkb} L.~Frachebourg, P.~L.~Krapivsky, and E.~Ben-Naim, 
              Phys.\ Rev.\ E {\bf 54}, 6186 (1996). 
              L.~Frachebourg, P.~L.~Krapivsky 
              J.\ Phys.\ A, {\bf 31}, L287 (1998). 

\bibitem{vN}  J.~von Neumann and O.~Morgenstern, 
              {\it Theory of Games and Economic Behavior}
              (Princeton University Press, Princeton, 1953).

\bibitem{cz}  D.~Challet and Y.~C.~Zhang, 
              Physica A {\bf 246}, 407 (1997).

\bibitem{cmz} D.~Challet, M.~Marsili, and R.~Zecchina, 
              Phys.\ Rev.\ Lett. {\bf 84}, 1824 (2000).
 
\bibitem{c}   A.~Cavagna, 
              Phys.\ Rev.\ E {\bf 59}, R3782 (1999). 
 
\bibitem{dr}  R.~D'Hulst and G.~J.~Rodgers, 
              Physica A {\bf 294}, 447 (2001).

\bibitem{col} See e.g. S.~Ulam, Adv.\ Appl.\ Math. {\bf 1}, 7 (1980); 
              E.~Ben-Naim and P.~L.~Krapivsky, 
              Phys.\ Rev.\ E {\bf 61}, R5 (2000);
              A.~Baldassarri, U.~Martini Bettolo Marconi, and A.~Puglisi,
              cond-mat/{\it 0105299}. 

\bibitem{rl}  P. Resibois and M. De Leener, 
              {\it Classical Kinetic Theory of Fluids} (Wiley, New York, 1977).

\end{references}
\end{document}